# Foundations for Relativistic Quantum Theory I:
# Feynman's Operator Calculus and the Dyson Conjectures

Tepper L. Gill[1,2,3] and W. W. Zachary[1,4]

[1]Department of Electrical Engineering
[2]Department of Mathematics
Howard University
Washington, DC 20059
E-mail: tgill@howard.edu

[3]Department of Physics
University of Michigan
Ann Arbor, Mich. 48109

[4]Department of Mathematics and Statistics
University of Maryland University College
College Park, Maryland 20742
E-mail: wwzachary@earthlink.net

**Abstract**

In this paper, we provide a representation theory for the Feynman operator calculus. This allows us to solve the general initial-value problem and construct the Dyson series. We show that the series is asymptotic, thus proving Dyson's second conjecture for QED. In addition, we show that the expansion may be considered exact to any finite order by producing the remainder term. This implies that every nonperturbative solution has a perturbative expansion. Using a physical analysis of information from experiment versus that implied by our models, we reformulate our theory as a sum over paths. This allows us to relate our theory to Feynman's path integral, and to prove Dyson's first conjecture that the divergences are in part due to a violation of Heisenberg's uncertainly relations.

## 1.0 Introduction

Following Dirac's quantization of the electromagnetic field in 1927[1], and his relativistic electron theory in 1928[2], the equations for quantum electrodynamics (QED) were developed by Heisenberg and Pauli[3,4] in the years 1929-30 (see Miller[5] and Schweber[6]). From the beginning, when researchers attempted to use the straightforward and physically intuitive time-dependent perturbation expansion to compute physical observerables, a number of divergent expressions appeared. Although it was known that the same problems also existed in classical electrodynamics, it was noted by Oppenheimer[7] that there was a fundamental difference in the quantum problem as compared to the classical one. (Dirac[9] had shown that, in the classical case, one could account for the problem of radiation reaction without directly dealing with the self-energy divergence by using both advanced and retarded fields and a particular limiting procedure.)

Early attempts to develop subtraction procedures for the divergent expressions were very discouraging because they depended on both the gauge and the Lorentz frame, making them appear ambiguous. Although the equations of QED were both Lorentz and gauge covariant, it was generally believed that, in a strict sense, they had no solutions expandable in powers of the charge. The thinking of the times was clearly expressed by Oppenheimer[8] in his 1948 report to the Solvay Conference, "If one wishes to explore these solutions, bearing in mind that certain infinite terms will, in a later theory, no longer be infinite, one needs a covariant way of identifying these terms; and for that, not merely the field equations themselves, but the whole method of approximation and solution must at all stages preserve covariance."

The solution to the problem posed by Oppenheimer was made (independently) by Tomonaga[10], Schwinger[11] and Feynman[12,13]. (These papers may be found in Schwinger[14].) Tomonaga introduced what is now known as the interaction representation and showed how the approximation process could be carried out in a covariant manner. Schwinger developed the general theory and applied it to many of the important problems. Feynman took a holistic view of physical reality in his development. He suggested that we view a physical event as occurring on a film which exposes more and more of the outcome as the film unfolds. His idea was to deal directly with the solutions to the equations describing the physical system, rather than the equations themselves. In addition to solving the problem posed by Oppenheimer, Feynman's approach led to a new perturbation series, which provided an easy, intuitive, and computationally simple method to study interacting particles while giving physical meaning to each term in his expansion.

Since Feynman's method and approach was so different, it was not clear how it related to that of Schwinger and Tomonaga. Dyson[15,16], made a major contribution. Dyson realized that Feynman and Schwinger were both dealing with different versions of Heisenberg's S-matrix. He then formally introduced time-ordering and provided a unified approach by demonstrating the equivalence of the Feynman and Schwinger-Tomonaga theories. This approach also allowed him to show how the Schwinger theory could be greatly simplified and extended to all orders of the perturbation expansion. Dyson's time-ordering idea was actually obtained from discussions with Feynman, who later explored and fully developed it into his time-ordered operator calculus[17].

## 1.1 Background

After the problem proposed by Oppenheimer was resolved, attitudes toward the renormalization program and quantum field theory could be classified into three basic groups. The first group consisted of those who were totally dissatisfied with the renormalization program. The second group considered the renormalization program an interim step and believed that the divergences were an indication of additional physics, which could not be reached by present formulations. The first two groups will not be extensively discussed in this paper. However, we can associate the names of Dirac and Landau with the first group, and Sakata and Schwinger with the second. (See Dirac[18], Sakata[19], Schwinger[20] and also Schweber[6].)

The third group was more positive, and directed its attention towards investigating the mathematical foundations of quantum field theory with the hope of providing a more orderly approach to the renormalization program (assuming that the theory proved consistent). This direction was clearly justified since part of the problem had been consistently blamed on a mathematical issue, the perturbation expansion. Indeed, the whole renormalization program critically depended on the expansion of the S-matrix in powers of the coupling constant. This concern was further supported since attempts to use the expansion when the coupling constant was large led to meaningless results. Additional unease could be attributed to the fact that, at that time, not much was actually known about the physically important cases where one was dealing with unbounded operator-valued functions (distributions).

Researchers working on the mathematical foundations of quantum electrodynamics, and quantum field theory adopted the name axiomatic field theory starting in the fifties. These researchers focused on trying to find out what could be learned about the existence of local relativistic quantum field theories based on certain natural assumptions which included the postulates of quantum mechanics, locality, Poincaré invariance, and a reasonable spectrum. This approach was initiated by the work of Wightman[21], and Lehmann, Symanzik and Zimmermann[22,23]. Here, the quantized field is interpreted mathematically as an operator-valued Schwartz distribution. Explicit use of the theory of distributions was a major step, which helped to partially make the theory (mathematically) sound by smoothing out the fields locally. (The recent paper by Wightman[24] provides an inspired introduction to the history of Heisenberg's early observations on the latter concept and its relationship to the divergences[25].)

The axiomatic approach proved very fruitful, providing the first rigorous proofs of a number of important general results, and attracted many able researchers. The favored name today is Algebraic Quantum Field Theory. The books by Jost[26], Streater and Wightman[27] and Bogolubov and Shirkov[28] are the classics, while more recent work can be found in Haag[29]. (See also the book by Bogolubov, Logunov and Todorov[30], and the recent review paper by Buchholz[31].)

For a number of reasons, most notably a lack of nontrivial examples, the axiomatic approach evolved in a number of directions. One major direction is called "constructive" quantum field theory. Here, one focuses on attempts to directly construct solutions of various model field theories, which either have exact (nonperturbative) solutions, or have an asymptotic perturbative expansion which can be summed to the exact solution. In this approach, instead of formulating the theory in Minkowski spacetime, one passes to imaginary time and formulates it in Euclidean space (an idea which first appeared in Dyson[15]). This leads to a formulation in terms of "Schwinger functions", also known as Euclidean Green's functions. The advantage of this approach is that hyperbolic equations are transformed to elliptic ones, and Gaussian kernels, for which a very rich set of analytic tools has been developed, replace Feynman kernels. The output of this enterprise is truly impressive. Constructive solutions have been obtained for a number of important models. Furthermore, this approach has given us a clearer picture of the problems associated with the rigorous construction of a relativistic quantum field theory and provided new mathematical methods. An early summary of this approach may be found in the lecture notes[32], while more recent progress is contained in the lecture notes[33], both edited by Velo and Wightman (see also references 41 and 6). The books by Glimm and Jaffe[34] and Simon[35] give a different flavor and point of departure.

Although a great deal of work has been done in constructive field theory over the last thirty years, many difficult problems still remain. For example, the appearance of difficulties with the constructive approach to polynomial types of field theories is discussed in the paper by Sokal[36]. He conjectured that the $\lambda\varphi^4_{\geq 4}$ theory ($\lambda\varphi^4$ in four or more spacetime dimensions) is a generalized free field, where $\lambda$ is the coupling constant. This theory represents a self-interacting boson field. The conjecture was proven by Aizenman and Graham[37] and Fröhlich[38]. Three years later, Gawedzki and Kupiainen[39] proved that, if we change the sign of the coupling constant, the solution exists (as a tempered distribution) and the perturbation expansion is asymptotic to the solution. This state of affairs led Wightman (reference 33, pg. 1) to lament that "We do not know whether the lack of an existence theorem for solutions with the "right" sign reflects the non-existence of solutions or merely the lack of a technique to construct them." Things are further complicated by the fact that the $\lambda\varphi_4^4$ theory has a perturbative solution! This led Gallavotti[45] to suggest that constructive approaches other than the ferromagnetic lattice approximation, used by Aizenman and Graham, and Fröhlich, may be required.

The most well known method for quantum field theory calculations is perturbative renormalization theory. This approach is discussed in most standard texts on quantum field theory and has an interesting history that is best told by Wightman[40]. (The first book to include Dyson's reformulation of the Feynman-Schwinger-Tomonaga theory is the classic by Jauch and Rohrlich[42].) Early work in the perturbative approach focused on the development of different renormalization methods with the hope of identifying those for which rigorous mathematical methods could be used. The methods generally consisted of two parts. First, the Green's functions were regularized in a relativistically and gauge invariant manner[28,40,41,46] to yield well-defined tempered distributions, even on the light cone. Then appropriate counter-terms were introduced so that, in the limit, when the regularization was removed, the various divergences of the S-matrix were also removed. It was found that all renormalization procedures are equivalent up to a finite renormalization (cf. references 40, 41). Today, theories are classified as "renormalizable" or "unrenormalizable" according as the number of renormalizable constants is finite or infinite, respectively.

Some model theories in less than four spacetime dimensions considered in constructive field theory belong to a special subclass of renormalizable theories called "super renormalizable", for which the renormalization process can be carried out without using perturbation theory[32-35]. For these theories, the renormalized perturbation series can be shown to be Borel summable to the exact nonperturbative solution. A nice summary of these developments was given by Glimm and Jaffe[34]. On the other hand, constructive models of the Gross-Neveu type are renormalizable but not super renormalizable (see reference 33).

Feldman et al[43] have studied the mathematical foundations of quantum electrodynamics from the perturbative point of view (see also Rosen in reference 33, pg. 201). Here, a renormalized formal power series (renormalized tree expansion) is obtained for a measure on the space of fields within the Euclidean formulation of QED. (The tree expansion method is an outgrowth of Wilson's[44] renormalization group approach as distilled by Gallavotti[45] and co-workers.) It is then shown that QED in four (Euclidean) dimensions is locally Borel summable. Their work is truly remarkable and represents the first (formal) proof that (Euclidean) quantum electrodynamics can be renormalized using gauge invariant counterterms. However, in general, it is a nontrivial problem to return from the Euclidean regime to Minkowski space. The return trip requires application of the Osterwalder-Schhrader reconstruction theorem (see reference 32). This theorem places conditions on the Euclidean Green's functions which guarantees analytic continuation back to the real-time vacuum expectation values. When these conditions are fulfilled, the Lehmann, Symanzik, and Zimmermann (LSZ)[22,23,32] reduction formulae may then be used to obtain the S-matrix. For technical reasons, they were not able to directly apply the Osterwalder-Schhrader theorem. They could still get back to QED in Minkowski spacetime by following the methods of Hepp[46] and Lowenstein and Speer[47]. However, nothing could be said about the convergence properties of their series.

## 1.2 Purpose

It is clear that Dyson's use of time-ordering was the fundamental conceptual tool which allowed him to relate the Feynman and Schwinger-Tomonaga theories. This tool has now become a natural part of almost every branch of physics and is even used in parts of engineering. Its importance to the foundations of quantum field theory led Segal[53] to suggest that the identification of mathematical meaning for Feynman's time-ordered operator calculus is one of the major problems. A number of investigators have attempted to solve this problem. Miranker and Weiss[54] showed how the Feynman ordering process could be done formally using the theory of Banach algebras. Nelson[55] used Banach algebras to developed a theory of "operants" as an alternate (formal) approach. Araki[56], motivated by the work of Fujiwara, used Banach algebras to develop yet another formal approach. (Fujiwara[57] had earlier suggested that the Feynman program could be implemented if one used a sheet of unit operators at every point except at time t, where the true operator should be placed.) Maslov[58] used the idea of a T-product to formally order operators and developed an operational theory. Another important approach to this problem via the idea of an index may be found in the works of Johnson and Lapidus[59-61], see also Johnson, Lapidus, and DeFacio[62].

This paper is a part of a new investigation into the physical and mathematical foundations of relativistic quantum theory. Our overall goal is to construct a self-consistent relativistic quantum theory of particles and fields. For this paper, we have two specific objectives. Our first (and major) objective is to construct a physically simple and computationally useful representation theory for the Feynman time-ordered operator calculus.

A correct formulation and representation theory for the Feynman time-ordered operator calculus should at least have the following desirable features:

1. It should provide a transparent generalization of current analytic methods without sacrificing the physically intuitive and computationally useful ideas of Feynman.

2. It should provide a clear approach to some of the mathematical problems of relativistic quantum theory.

3. It should explain the connection with path integrals.

In the course of his analysis, unification, and simplification of the Feynman-Schwinger-Tomonaga theory, Dyson made two important suggestions (conjectures). The first conjecture concerned the divergences in QED, while the second was concerned with the convergence of the renormalized perturbation series. In addressing the problem of divergences, Dyson conjectured that they may be due to an idealized conception of measurability resulting from the infinitely precise knowledge of the space-time positions of particles (implied by our Hamiltonian formulation) which leads to a violation of the Heisenberg uncertainty principle. This point of view can be traced directly to the Bohr-Rosenfeld theory of measurability for field operators and, according to Schweber[6], is an outgrowth of Dyson's discussions with Oppenheimer.

In addressing the renormalized S-matrix[16], Dyson suggested that it might be more reasonable to expect the expansion to be asymptotic rather than convergent and gave physical arguments to support his claim. The lack of a clear mathematical framework made it impossible to formulate and investigate his suggestions.

Schweber[6] notes that Dyson made two other well-known conjectures. The "overlapping divergences" conjecture was proved by Salam[48], Ward[49], Mills and Yang[50], and Hepp[51]. Dyson's conjecture that a certain Feynman integral converges, necessary for showing that the ultraviolet divergences cancel to all orders, was proved by Weinberg[52].

Our second objective is to provide proofs of the above two conjectures under general conditions that should apply to any formulation of quantum field theory which does not abandon Hamiltonian generators for unitary solution operators. The proof of the first conjecture is, to some extent expected, and is a partial vindication of our belief in the consistency of quantum electrodynamics in the sense that the ultraviolet problem is caused by an effect that is basically "simple". Such a result is partly anticipated since the effect can be made to disappear via appropriate cutoffs. We also identify (special) conditions under which the renormalized perturbation series may actually converge. A proof of the above conjectures is implicit in, and is one of the major achievements of constructive field theory for the models studied. In fact, these theories verify a stronger version of the second conjecture since, as noted earlier, the renormalized perturbation series is summable to the true solution.

**1.3 Summary**

The work in this paper is both a generalization and simplification of earlier work[63-65] that is easier and requires the weakest known conditions. We construct a new representation Hilbert space and von Neumann algebra for the Feynman (time-ordered) operator calculus. In order to make the theory applicable to other areas, we develop it using semigroups of contractions and the Riemann integral. A contraction semigroup on a Hilbert space $\mathcal{H}$ can always be extended to a unitary group on a larger space $\mathcal{H}'$. Thus, for quantum theory we may replace the semigroups by unitary groups and assume that our space is $\mathcal{H}'$ without any loss in understanding.

The Riemann integral can be easily replaced by the operator-valued Riemann-complete integral of Henstock[66] and Kurzweil[67], which generalizes the Bochner and Pettis integrals (see Gill[63]). This integral is easier to understand (and learn) compared to the Lebesgue or Bochner integrals, and provides useful variants of the same theorems that have made those integrals so important. Furthermore, it arises from a simple (transparent) generalization of the Riemann integral that was taught in elementary calculus. Its usefulness in the construction of Feynman path integrals was first shown by Henstock[68], and has been further explored in the recent book by Muldowney[69].

In Section 1.4 we provide a brief review of the necessary operator theory in order to make the paper self-contained. In Section 2 we construct an infinite tensor product Hilbert space and define what we mean by time ordering. In Section 3 we construct time-ordered integrals and evolution operators and prove that they have the expected properties. In Section 4 we define what is meant by the phase "asymptotic in the sense of Poincaré" for operators, and use it to prove Dyson's second conjecture for contraction semigroups. We then discuss conditions under which the perturbation series may be expected to converge.

In Section 5 we take a photograph of a track left by an elementary particle in a bubble chamber as a prototype to conduct a physical analysis of what is actually known from experiment. This approach is used to rederive our time-ordered evolution operator as the limit of a probabilistic sum over paths. We use it to briefly discuss our theory in relationship to the Feynman path integral, and show that it provides a general and natural definition for the path integral that is independent of measure theory and the space of continuous paths.

The results from Section 5 are applied to the S-matrix expansion in Section 6 to provide a formulation and proof of Dyson's first conjecture. In particular, we show that, within our formulation, the assumption of precise time information over a particle's trajectory introduces an infinite amount of energy into the system at each point in time. We use Dyson's original notation partly for reasons of nostalgia, but also to point out what we are not able to explain within our framework. Also, since all renormalization procedures are equivalent, there is no loss.

**1.4 Operator Theory**

In this subsection we establish notation and quote some results from operator theory used in the paper. Let $\mathcal{H}$ denote a separable Hilbert space over **C** (complex numbers), **B**($\mathcal{H}$) the set of bounded linear operators, and **C**($\mathcal{H}$) the set of closed densely defined linear operators on $\mathcal{H}$.

**Definition 1.0** A family of bounded linear operators $\{U(t,0), 0 \leq t < \infty\}$ defined on $\mathcal{H}$ is a *strongly continuous semigroup (or $C_0$- semigroup)* if

1. $U(0,0) = I$,  2. $U(t+s, 0) = U(t,0)U(s,0)$,  3. $\lim_{t \to 0} U(t,0)\varphi = \varphi$, $\forall \varphi \in \mathcal{H}$

$U(t,0)$ is a contraction semigroup in case $\|U(t,0)\| \leq 1$. If we replace 2 by

2'. $U(t,\tau) = U(t,s)U(s,\tau)$, $0 \leq \tau \leq s \leq t < \infty$, then we call $U(t,\tau)$ a *strongly continuous evolution family.*

**Definition 1.2** A densely defined operator $H$ is said to be *maximal dissipative* if $\operatorname{Re}\langle H\varphi, \varphi \rangle \leq 0$, $\forall \varphi \in D(H)$, and $\mathcal{R}\operatorname{an}(I - H) = \mathcal{H}$ (range of $(I - H)$).

The following results may be found in Goldstein[70] or Pazy[71].

**Theorem 1.2** *Let $U(t,0)$ be a $C_0$-semigroup of contraction operators on $\mathcal{H}$. Then*

1) $H\varphi = \lim_{t \to 0} \dfrac{U(t,0)\varphi - \varphi}{t}$ *exists for $\varphi$ in a dense set.*

2) $R(z, H) = (zI - H)^{-1}$ *exists for $z > 0$ and $\|R(z, H)\| \leq \dfrac{1}{z}$.*

**Theorem 1.3** *Suppose $H$ is a maximal dissipative operator. Then $H$ generates a unique $C_0$-semigroup $\{U(t,0) \mid 0 \leq t < \infty\}$ of contraction operators on $\mathcal{H}$.*

**Theorem 1.4** *If $H$ is densely defined with both $H$ and $H^*$ dissipative, then $H$ is maximal dissipative.*

**2.0 Infinite Tensor Product von Neumann Algebras**

In this section we define time-ordered operators and construct the representation space which will be used in the next section to develop our theory of time-ordered integrals and evolution operators. Much of the material in this section was developed by von Neumann[72] for other purposes, but is perfectly suited for our program. In order to see how natural our approach is, let $\mathcal{H}_\otimes = \hat{\otimes}_s \mathcal{H}(s)$ denote the infinite tensor product Hilbert space of von Neumann, where $\mathcal{H}(s) = \mathcal{H}$ for s ∈ [a,b] and $\hat{\otimes}$ denotes closure. If $\mathbf{B}(\mathcal{H}_\otimes)$ is the set of bounded operators on $\mathcal{H}_\otimes$, define $\mathbf{B}(\mathcal{H}(t)) \subset \mathbf{B}(\mathcal{H}_\otimes)$ by

$$\mathbf{B}(\mathcal{H}(t)) = \{\mathbf{H}(t) | \ \mathbf{H}(t) = \hat{\otimes}_{a \geq s > t} I_s \otimes H(t) \otimes (\otimes_{t > s \geq -a} I_s), \forall H(t) \in \mathbf{B}(\mathcal{H})\}, \quad (2.1a)$$

where $I_s$ denotes an identity operator, and let $\mathbf{B}^\#(\mathcal{H}_\otimes)$ be the uniform closure of the von Neumann algebra generated by the family $\{\mathbf{B}(\mathcal{H}(t)), | t \in E\}$. If the family $\{H(t) | t \in E\}$ is in $\mathbf{B}(\mathcal{H})$, then the corresponding operators $\{\mathbf{H}(t) | t \in E\} \in \mathbf{B}^\#(\mathcal{H}_\otimes)$ commute when acting at different times: $t \neq s \Rightarrow$

$$\mathbf{H}(t)\mathbf{H}(s) = \mathbf{H}(s)\mathbf{H}(t). \quad (2.1b)$$

**Definition 2.0** The smallest space $\mathcal{FD}_\otimes \subseteq \mathcal{H}_\otimes$ which, leaves the family $\{\mathbf{H}(t) \ | t \in E\}$ invariant is called a <u>Feynman-Dyson space</u> for the family. (This is the film.)

We need the following results about operators on $\mathcal{H}_\otimes$.

**Theorem 2.1:** (von Neumann[72] *The mapping $\mathbf{T}_\theta^t$: $\mathbf{B}(\mathcal{H}) \to \mathbf{B}(\mathcal{H}(t))$ is an isometric isomorphism of algebras. (We call $\mathbf{T}_\theta^t$ <u>the time-ordering morphism</u>.)*

**Definition 2.2** *The vector $\Phi = \otimes_s \phi_s$ is said to be equivalent to $\Psi = \otimes_s \psi_s$ and we write $\Phi \approx \Psi$, if and only if*

$$\sum_s \left| \langle \phi_s, \psi_s \rangle_s - 1 \right| < \infty. \qquad (2.2)$$

Here, $\langle \cdot, \cdot \rangle_s$ is the inner product on $\mathcal{H}(s)$, and it is understood that the sum is meaningful only if at most a countable number of terms are different from zero.

Let $\mathcal{H}_\Phi = cl\left\{ \Psi \mid \Psi = \sum_{i=1}^n \Psi_i, \Psi_i \approx \Phi, n \in \mathbf{N} \right\}$ (closure), $\Phi \in \mathcal{H}_\otimes$, and let $\mathbf{P}_\Phi$ denote the projection from $\mathcal{H}_\otimes$ onto $\mathcal{H}_\Phi$. The space $\mathcal{H}_\Phi$ is known as the *incomplete tensor product generated by* $\Phi$. The details on incomplete tensor product spaces as well as proofs of the next two theorems may be found in von Neumann[72].

**Theorem 2.3** *The relation defined above is an equivalence relation on $\mathcal{H}_\otimes$ and*

1) *if $\Psi$ is not equivalent to $\Phi$, then $\mathcal{H}_\Phi \cap \mathcal{H}_\Psi = \{0\}$ (i.e., $\mathcal{H}_\Phi \perp \mathcal{H}_\Psi$);*

2) *if $\psi_s \neq \phi_s$ occurs for at most a finite number of s, then $\Phi = \otimes_s \phi_s \approx \Psi = \otimes_s \psi_s$;*

3) *if $\mathbf{T} \in \mathbf{B}^\#(\mathcal{H}_\otimes)$, then $\mathbf{P}_\Phi \mathbf{T} = \mathbf{T} \mathbf{P}_\Phi$ so that $\mathbf{P}_\Phi \mathbf{T} \in \mathbf{B}^\#(\mathcal{H}_\Phi)$.*

The second condition in Theorem 2.3 implies that, for each fixed $\Phi = \otimes_s \phi_s$, there is an uncountable number of $\Psi = \otimes_s \psi_s$ equivalent to $\Phi$, while the third condition implies that every bounded linear operator on $\mathcal{H}_\otimes$ restricts to a bounded linear operator on $\mathcal{H}_\Phi$ for each $\Phi$.

We can now construct our film $\mathcal{FD}_\otimes$. Let $\{e^i | i \in N\}$ denote an arbitrary ordered complete orthonormal basis (c.o.b) for $\mathcal{H}$. For each $t \in \mathrm{E}, i \in N$, let $e_t^i = e^i$, $E^i = \otimes_{t \in \mathrm{E}} e_t^i$, and define $\mathcal{FD}^i$ to be the incomplete tensor product generated by the vector $E^i$. Setting $\mathcal{FD}_\otimes = \bigoplus_{i=1}^\infty \mathcal{FD}^i$, it will be clear in the next section that $\mathcal{FD}_\otimes$ is (one of an infinite number of) the natural representation space(s) for Feynman's time-ordered operator theory. It should be noted that $\mathcal{FD}_\otimes$ is a nonseparable Hilbert (space) bundle over [a, b]. However, it is not hard to see that each fiber is isomorphic to $\mathcal{H}$.

In order to facilitate the proofs in the next section, we need an explicit basis for each $\mathcal{FD}^i$. To construct it, fix i and let $f^i$ denote the set of all functions $\{j(t)| t \in E\}$ mapping $E \to N \cup \{0\}$ such that $j(t)$ is zero for all but a finite number of $t$. Let $I(j) = \{j(t)| t \in E\}$ denote the function j and set $E^i_{I(j)} = \otimes_{t \in E} e^i_{t,j(t)}$ with $e^i_{t,0} = e^i$, and $j(t) = k \Rightarrow e^i_{t,k} = e^k$.

**Theorem 2.4** *The set* $\{E^i_{I(j)}|I(j) \in f^i\}$ *is a (c.o.b) for each* $\mathcal{FD}^i$.

For each $\Phi^i, \Psi^i \in F^i$, set $a^i_{I(j)} = \langle \Phi^i, E^i_{I(j)} \rangle$, $b^i_{I(j)} = \langle \Psi^i, E^i_{I(j)} \rangle$, so that $\Phi^i = \sum_{I(j) \in F^i} a^i_{I(j)} E^i_{I(j)}$, $\Psi^i = \sum_{I(j) \in F^i} b^i_{I(j)} E^i_{I(j)}$ and $\langle \Phi^i, \Psi^i \rangle = \sum_{I(j) \in F^i} a^i_{I(j)} \bar{b}^i_{I(k)} \langle E^i_{I(j)}, E^i_{I(k)} \rangle$. Now,

$\langle E^i_{I(j)}, E^i_{I(k)} \rangle = \prod_t \langle e^i_{t,I(j)}, e^i_{t,I(k)} \rangle = 0$, unless $j(t) = k(t), \forall t \in E$, so that
$\langle \Phi^i, \Psi^i \rangle = \sum_{I(j) \in F^i} a^i_{I(j)} \bar{b}^i_{I(j)}$.

We need the notion of an exchange operator. (Theorem 2.6 is in reference 63.)

**Definition 2.5** *An exchange operator* $\mathbf{E}[t,t']$ *is a linear map defined for pairs $t, t' \in [a, b]$ such that:*

1. $\mathbf{E}[t,t']: \mathbf{B}(\mathcal{H}(t)) \to \mathbf{B}(\mathcal{H}(t'))$ *onto*,
2. $\mathbf{E}[t,s]\mathbf{E}[s,t'] = \mathbf{E}[t,t']$,
3. $\mathbf{E}[t,t']\mathbf{E}[t',t] = 1$,
4. *if* $s \neq t, t'$, *then* $\mathbf{E}[t,t']\mathbf{H}(s) = \mathbf{H}(s) \ \forall \mathbf{H}(s) \in \mathbf{B}(\mathcal{H}(s))$.

**Theorem 2.6**

1) $\mathbf{E}[\cdot,\cdot]$ exists and is a Banach algebra isomorphism on $\mathbf{B}^\#(\mathcal{H}_\otimes)$.

2) $\mathbf{E}[s,s']\mathbf{E}[t,t'] = \mathbf{E}[t,t']\mathbf{E}[s,s']$ for distinct pairs $(s,s')$ and $(t,t')$ in E.

## 3.0 Time-Ordered Integrals

In this section we construct time-ordered integrals and evolution operators for a fixed family $\{H(t)|\, t \in \mathrm{E}\} \subset C(\mathcal{H})$ of generators of contraction semigroups on $\mathcal{H}$. We assume that, for each t, $H(t)$ and $H^*(t)$ are dissipative (so that the family is maximal dissipative for each t). In the following discussion we adopt the notation:

1). (e.o.v): "except for at most one s value";

2). (e.f.n.v): "except for an at most finite number of s values"; and

3). (a.s.c): "almost surely and the exceptional set is at most countable".

The s value referred to is in our fixed interval E.

For the given family $\{H(t)|\, t \in \mathrm{E}\} \subset C(\mathcal{H})$, define $\exp\{\tau \mathbf{H}(t)\}$ by

$$\exp\{\tau \mathbf{H}(t)\} = \hat{\bigotimes}_{s \in [b,t)} \mathbf{I}_s \otimes \left(\exp\{\tau H(t)\}\right) \otimes \left(\bigotimes_{s \in (t,a]} \mathbf{I}_s\right), \qquad (3.1)$$

and set $\mathbf{H}_z(t) = z\mathbf{H}(t)\mathbf{R}(z,\mathbf{H}(t))$, $z > 0$, where $\mathbf{R}(z,\mathbf{H}(t)) = (z\mathbf{I}_\otimes - \mathbf{H}(t))^{-1}$ is the resolvent of $\mathbf{H}(t)$. It is known that $H_z(t)$ generates a uniformly bounded contraction semigroup and $\lim_{z \to \infty} H_z(t)\phi = H(t)\phi$ for $\phi \in D(H(\mathrm{t}))$.

**Theorem 3.1** *Suppose for each t, $\{H(t)|t \in \mathrm{E}\} \subset C(\mathcal{H})$ generates a strongly continuous contraction semigroup on $\mathcal{H}$. Then $\mathbf{H}(t)\mathbf{H}_z(t)\Phi = \mathbf{H}_z(t)\mathbf{H}(t)\Phi$, $\Phi \in \mathrm{D}$, (where* D *denotes the domain of the family* $\{\mathbf{H}(t)|t \in \mathrm{E}\}$), *and*

1. *The family $\{\mathbf{H}_z(t)|t\in E\}$ generates a uniformly bounded contraction semigroup on $\mathcal{FD}_\otimes$ for each t and $\lim_{z\to\infty}\mathbf{H}_z(t)\Phi = \mathbf{H}(t)\Phi$, $\Phi \in D$.*

2. *The family $\{\mathbf{H}(t)|t\in E\} \subset \mathbf{C}(\mathcal{H}_\otimes)$ generates a strongly continuous contraction semigroup on $\mathcal{FD}_\otimes$ (so that $\{\mathbf{H}(t)|t\in E\} \subset \mathbf{C}(\mathcal{FD}_\otimes)$).*

**Proof:** The proof of 1. is standard. Note that $\mathbf{H}_z(t) = z^2\mathbf{R}(z,\mathbf{H}(t)) - z\mathbf{I}_\otimes$ and $\|\mathbf{R}(z,\mathbf{H}(t))\|_\otimes \leq 1/z$, so $\|\exp\{s\mathbf{H}_z(t)\}\|_\otimes = \|\exp\{-sz\}\exp\{sz^2\mathbf{R}(z,\mathbf{H}(t))\}\|_\otimes \leq 1$. Now recall that $\lim_{z\to\infty}\{z\mathbf{R}(z,\mathbf{H}(t))\Phi\} = \Phi$, $\Phi \in \mathcal{FD}_\otimes$, so that, for $\Phi \in D$, we have that

$$\lim_{z\to\infty}\mathbf{H}_z(t)\Phi = \lim_{z\to\infty}\{z\mathbf{H}(t)\mathbf{R}(z,\mathbf{H}(t))\Phi\} = \lim_{z\to\infty}\{z\mathbf{R}(z,\mathbf{H}(t))\}\mathbf{H}(t)\Phi = \mathbf{H}(t)\Phi.$$

To prove 2., first recall (Gill[73]) that a tensor product norm, $\|\cdot\|_\otimes$, is uniform if, for $\hat{\otimes}_{s\in E}T_s \in \mathbf{B}(\mathcal{H}_\otimes)$,

$$\left\|\hat{\otimes}_{s\in E} T_s\right\|_\otimes \leq \prod_{s\in E}\|T_s\|. \qquad (3.2)$$

Using the uniform property of the (Hilbert space) tensor product norm, it is easy to see that $\exp\{\tau\mathbf{H}(t)\}$ is a contraction semigroup.

To prove strong continuity, we need to identify a dense core for the family $\{\mathbf{H}(t)|t\in E\} \subset \mathbf{C}(\mathcal{FD}_\otimes)$. Let $D_1$ denote the ordered tensor product of the domains of the family $\{H(t)| t \in E\} \subset \mathbf{C}(\mathcal{H})$, (so that $D_1 \subset D$)

$$D_1 = \otimes_{s\in E} D(H(s)) = \left\{\sum_{i=1}^n \otimes_s \varphi_s^i \Big| \varphi_s^i \in D(H(s)), s \in E\right\}. \qquad (3.3)$$

It is clear that $D_1$ is a dense core in $\mathcal{H}_\otimes$, so $D_0 = D_1 \cap \mathcal{FD}_\otimes$ is a dense core in $\mathcal{FD}_\otimes$. Using our standard basis, if $\Phi, \Psi \in D_0$, $\Phi = \sum_i \sum_{I(j)} a_{I(j)}^i E_{I(j)}^i$, $\Psi = \sum_i \sum_{I(k)} b_{I(k)}^i E_{I(k)}^i$;

then, since $(\exp\{\tau\mathbf{H}(t)\}-\mathrm{I}_\otimes)$ is invariant on $\mathcal{FD}^i$ and $\mathrm{I}_\otimes$ is the identify on $\mathcal{FD}_\otimes$, we have

$$\langle(\exp\{\tau\mathbf{H}(t)\}-\mathrm{I}_\otimes)\Phi,\Psi\rangle = \sum_i \sum_{I(j)} \sum_{I(k)} a^i_{I(j)} \bar{b}^i_{I(k)} \langle(\exp\{\tau\mathbf{H}(t)\}-\mathrm{I}_\otimes)E^i_{I(j)}, E^i_{I(k)}\rangle, \quad (3.4a)$$

and

$$\langle(\exp\{\tau\mathbf{H}(t)\}-\mathrm{I}_\otimes)E^i_{I(j)}, E^i_{I(k)}\rangle = \prod_{s\neq t} \langle e^i_{s,j(s)}, e^i_{s,k(s)}\rangle \langle(\exp\{\tau H(t)\}-\mathrm{I})e^i_{t,j(t)}, e^i_{t,k(t)}\rangle \quad (3.4b)$$

$$= \langle(\exp\{\tau H(t)\}-\mathrm{I})e^i_{t,j(t)}, e^i_{t,j(t)}\rangle \text{ (e.o.v)},$$
$$= \langle(\exp\{\tau H(t)\}-\mathrm{I})e^i, e^i\rangle \text{ (e.f.n.v.)},$$

$$\Rightarrow \langle(\exp\{\tau\mathbf{H}(t)\}-\mathrm{I}_\otimes)\Phi,\Psi\rangle = \sum_i \sum_{I(j)} a^i_{I(j)} \bar{b}^i_{I(j)} \langle(\exp\{\tau H(t)\}-\mathrm{I})e^i, e^i\rangle \text{ (a.s.c)}. \quad (3.4c)$$

Since all sums are finite, we have

$$\lim_{\tau \to 0}\langle(\exp\{\tau\mathbf{H}(t)\}-\mathrm{I}_\otimes)\Phi,\Psi\rangle = \sum_i \sum_{I(j)} a^i_{I(j)} \bar{b}^i_{I(j)} \left\{\lim_{\tau \to 0}\langle(\exp\{\tau H(t)\}-\mathrm{I})e^i, e^i\rangle\right\} = 0 \text{ (a.s.c)}. (3.4d)$$

The if and only if part is now clear. Since $\exp\{\tau\mathbf{H}(t)\}$ is bounded on $\mathcal{H}_\otimes$ and the above limit exists on $\mathrm{D}_0$ (which is dense in $\mathcal{FD}_\otimes$), we see that $\exp\{\tau\mathbf{H}(t)\}$ extends to a contraction semigroup on $\mathcal{FD}_\otimes$. Now use the fact that, if a bounded semigroup converges weakly to the identity, it converges strongly (see Pazy[71], pg. 44).

We now assume that the family $\{H(t)|\ t\in \mathrm{E}\} \subset \mathbf{C}(\mathcal{H})$ has a weak Riemann integral $Q = \int_a^b H(t)dt \in \mathbf{C}(\mathcal{H})$. It follows that the family $\{H_z(t)|\ t\in \mathrm{E}\} \subset \mathbf{B}(\mathcal{H})$ also has a weak Riemann integral $Q_z = \int_a^b H_z(t)dt \in \mathbf{B}(\mathcal{H})$. Let $\mathrm{P}_n$ be a sequence of partitions (of E) so that the mesh $\mu(P_n) \to 0$ as $n \to \infty$. Set

$$Q_{z,n} = \sum_{l=1}^{n} H_z(\bar{t}_l)\Delta t_l, \; Q_{z,m} = \sum_{q=1}^{m} H_z(\bar{s}_q)\Delta s_q; \; \mathbf{Q}_{z,n} = \sum_{l=1}^{n} \mathbf{H}_z(\bar{t}_l)\Delta t_l, \; \mathbf{Q}_{z,m} = \sum_{q=1}^{m} \mathbf{H}_z(\bar{s}_q)\Delta s_q; \text{ and}$$

$$\Delta Q_z = Q_{z,n} - Q_{z,m}, \; \Delta \mathbf{Q}_z = \mathbf{Q}_{z,n} - \mathbf{Q}_{z,m}. \text{ Let } \Phi, \Psi \in D_0; \; \Phi = \sum_i \Phi^i = \sum_i^J \sum_{I(j)}^K a_{I(j)}^i E_{I(j)}^i,$$

$$\Psi = \sum_i^L \Psi^i = \sum_i^L \sum_{I(k)}^M b_{I(k)}^i E_{I(k)}^i, \text{ and set } \phi = \sum_i^J \sum_{I(j)}^K a_{I(j)}^i e^i \text{ and } \psi = \sum_i^J \sum_{I(j)}^K b_{I(j)}^i e^i. \text{ Then we}$$

have:

**Theorem 3.2** (First Fundamental Theorem for Time-Ordered Integrals)

$$\langle \Delta \mathbf{Q}_z \Phi, \Psi \rangle = \sum_i^J \sum_{I(j)}^K a_{I(j)}^i \bar{b}_{I(j)}^i \langle \Delta Q_z e^i, e^i \rangle \text{ (a.s.c.)}. \tag{3.5}$$

**Note** The form of (3.5) is quite general since $\Delta \mathbf{Q}_z$ can be replaced by other terms which also give a true relationship. For example, it is easy to show that the family $\{\mathbf{H}_z(t) | t \in E\}$ is weakly measurable, weakly continuous, weakly differentiable, etc if and only if the same is true for the family $\{H_z(t) | t \in E\}$.

**Proof:** $\langle \Delta \mathbf{Q}_z \Phi, \Psi \rangle = \sum_i \sum_{I(j)} \sum_{I(k)} a_{I(j)}^i \bar{b}_{I(k)}^i \langle \Delta \mathbf{Q}_z E_{I(j)}^i, E_{I(k)}^i \rangle$ (we omit the upper limit). Now

$$\langle \Delta \mathbf{Q}_z E_{I(j)}^i, E_{I(k)}^i \rangle = \sum_{l=1}^{n} \Delta t_l \langle \mathbf{H}_z(\bar{t}_l) E_{I(j)}^i, E_{I(k)}^i \rangle - \sum_{q=1}^{m} \Delta s_q \langle \mathbf{H}_z(\bar{s}_q) E_{I(j)}^i, E_{I(k)}^i \rangle$$

$$= \sum_{l=1}^{n} \Delta t_l \prod_{t \neq \bar{t}_l} \langle e_{t,j(t)}^i, e_{t,k(t)}^i \rangle \langle H_z(\bar{t}_l) e_{\bar{t}_l, j(\bar{t}_l)}^i, e_{\bar{t}_l, k(\bar{t}_l)}^i \rangle - \sum_{q=1}^{m} \Delta s_q \prod_{t \neq \bar{s}_q} \langle e_{t,j(t)}^i, e_{t,k(t)}^i \rangle \langle H_z(\bar{s}_q) e_{\bar{s}_q, j(\bar{s}_q)}^i, e_{\bar{s}_q, k(\bar{s}_q)}^i \rangle$$

$$= \sum_{l=1}^{n} \Delta t_l \langle H_z(\bar{t}_l) e_{\bar{t}_l, j(\bar{t}_l)}^i, e_{\bar{t}_l, j(\bar{t}_l)}^i \rangle - \sum_{q=1}^{m} \Delta s_q \langle H_z(\bar{s}_q) e_{\bar{s}_q, j(\bar{s}_q)}^i, e_{\bar{s}_q, j(\bar{s}_q)}^i \rangle$$

$$= \langle \Delta Q_z e^i, e^i \rangle \text{ (e.f.n.v)}. \text{ This result leads to (3.5).}$$

**Theorem 3.3** (Second Fundamental Theorem for Time-Ordered Integrals)
*If the family $\{H_z(t) | t \in E\}$ has a weak Riemann (Riemann-Complete) integral, then*

1. *the family* $\{\mathbf{H}_z(t) | t \in E\} \subset \mathbf{B}^{\#}(\mathcal{FD}_{\otimes})$ *has a weak Riemann (Riemann-Complete) integral.*

2. *If, in addition, we assume that for each* $\Phi$ *with* $\|\Phi\| = 1$,

$$\sup_{t \in E}\left|\int_a^t \left(\|\mathbf{H}_z(s)\Phi\|^2 - |\langle \mathbf{H}_z(s)\Phi, \Phi\rangle|^2\right)ds\right| < \infty, \qquad (3.6)$$

*then the family* $\{\mathbf{H}_z(t) | t \in E\}$ *has a strong integral* $\mathbf{Q}_z[t,a] = \int_a^t \mathbf{H}_z(s)ds$ *which generates a uniformly continuous contraction semigroup on* $\mathcal{FD}_{\otimes}$.

**Notes:**

**1.** It is sufficient that $\sup_{t \in E}\left|\int_a^t \left(\|\mathbf{H}_z(s)E^i\|^2 - |\langle \mathbf{H}_z(s)E^i, E^i\rangle|^2\right)ds\right| < \infty$ for each i.

**2.** Condition (3.6) is satisfied if $\|\mathbf{H}_z(s)E^i\|^2$ is Lebesgue integrable for each i. In this case, we replace the Riemann integral by the Riemann-Complete integral.

**3.** In general, the family $\{\mathbf{H}_z(t) | t \in E\}$ need not be a Bochner or Pettis integral, as it is not required that $\|\mathbf{H}_z(t)\Phi\|, \langle \mathbf{H}_z(t)\Phi, \Phi\rangle$ be (square) Lebesgue integrable. It is possible that $\int_a^b \|\mathbf{H}_z(t)\Phi\|^2 dt = \infty$ & $\int_a^b |\langle \mathbf{H}_z(t)\Phi, \Phi\rangle|^2 dt = \infty$, while (3.6) is zero.

For example, let $f(t)$ be any nonabsolutely (square) integrable function and set $\mathbf{H}_z(t) = f(t)I_{\otimes}$. Then the above possibility holds while $\int_a^t \left(\|\mathbf{H}_z(s)\Phi\|^2 - |\langle \mathbf{H}_z(s)\Phi, \Phi\rangle|^2\right)ds \equiv 0$ for all $t$ in E.

**Proof:** The proof of 1. is easy and follows from (3.5). To see that (3.6) makes $\mathbf{Q}_z$ a strong limit, let $\Phi \in D_0$. Then

$$\langle \mathbf{Q}_{z,n}\Phi, \mathbf{Q}_{z,n}\Phi \rangle = \sum_{i}\sum_{I(j),I(h)}^{J\ K} a_{I(j)}^{i}\overline{a}_{I(h)}^{i} \left( \sum_{k,m}^{n} \Delta t_{k}\Delta t_{m} \langle H_{z}(s_{k})E_{I(j)}^{i}, H_{z}(s_{m})E_{I(h)}^{i} \rangle \right)$$

$$= \sum_{i}\sum_{I(j)}^{J\ K} |a_{I(j)}^{i}|^{2} \left( \sum_{k\neq m}^{n} \Delta t_{k}\Delta t_{m} \langle H_{z}(s_{k})e_{s_{k},j(s_{k})}^{i}, e_{s_{k},j(s_{k})}^{i} \rangle \langle e_{s_{m},j(s_{m})}^{i}, H_{z}(s_{m})e_{s_{m},j(s_{k})}^{i} \rangle \right)$$

$$+ \sum_{i}\sum_{I(j)}^{J\ K} |a_{I(j)}^{i}|^{2} \left( \sum_{k}^{n} (\Delta t_{k})^{2} \langle H_{z}(s_{k})e_{s_{k},j(s_{k})}^{i}, H_{z}(s_{k})e_{s_{k},j(s_{k})}^{i} \rangle \right). \quad (3.7)$$

This can be rewritten as

$$\|\mathbf{Q}_{z,n}\Phi\|_{\otimes}^{2} = \sum_{i}\sum_{I(j))}^{J\ K} |a_{I(j)}^{i}|^{2} \Big\{ |\langle Q_{z,n}e^{i}, e^{i}\rangle|^{2}$$
$$+ \sum_{k}^{n} (\Delta t_{k})^{2} \Big( \|H_{z}(s_{k})e^{i}\|^{2} - |\langle H_{z}(s_{k})e^{i}, e^{i}\rangle|^{2} \Big) \Big\}, (a.s.c). \quad (3.8)$$

The last term can be written as

$$\left| \sum_{k,}^{n} (\Delta t_{k})^{2} \Big( \|H_{z}(s_{k})e^{i}\|^{2} - |\langle H_{z}(s_{k})e^{i}, e^{i}\rangle|^{2} \Big) \right| \le \mu_{n} M \sup_{t\in E} \left| \int_{a}^{t} \Big( \|H_{z}(s)e^{i}\|^{2} - |\langle H_{z}(s)e^{i}, e^{i}\rangle|^{2} \Big) ds \right|,$$

where M is a constant and $\mu_{n}$ is the mesh of $P_{n}$, with $\mu_{n} \to 0$ as $n \to \infty$. Now note that $\|\mathbf{H}_{z}(t)E^{i}\|_{\otimes} = \|H_{z}(t)e^{i}\|$ and $\langle \mathbf{H}_{z}(t)E^{i}, E^{i}\rangle = \langle H_{z}(t)e^{i}, e^{i}\rangle$ (e.o.v) so that

$$\sup_{t\in E}\left|\int_{a}^{t}\Big(\|H_{z}(s)e^{i}\|^{2} - |\langle H_{z}(s)e^{i}, e^{i}\rangle|^{2}\Big)ds\right| = \sup_{t\in E}\left|\int_{a}^{t}\Big(\|\mathbf{H}_{z}(s)E^{i}\|^{2} - |\langle \mathbf{H}_{z}(s)E^{i}, E^{i}\rangle|^{2}\Big)ds\right| \text{ (a.s.c)}.$$

We can now use (3.6) to get

$$\|\mathbf{Q}_{z,n}\Phi\|_{\otimes}^{2} \le \sum_{i}\sum_{I(j))}^{J\ K} |a_{I(j)}^{i}|^{2} \Big\{ |\langle Q_{z,n}e^{i}, e^{i}\rangle|^{2} + \mu_{n} M \sup_{t}\left|\int_{a}^{t}\Big(\|\mathbf{H}_{z}(t)E^{i}\|^{2} - |\langle \mathbf{H}_{z}(t)E^{i}, E^{i}\rangle|^{2}\Big)ds\right| \Big\}, (a.s.c).$$

Thus, $\mathbf{Q}_{z,n}\Phi$ converges strongly to $\mathbf{Q}_z\Phi$ on $D_0$ and hence has a strong limit on $\mathcal{FD}_\otimes$. To show that $Q_z[t,a]$ generates a uniformly continuous contraction, it suffices to show that $Q_z[t,a]$ and $Q_z^*[t,a]$ are dissipative. Let $\Phi$ be in $D_0$, then
$$\langle \mathbf{Q}_z[t,a]\Phi, \Phi \rangle = \sum_i^J \sum_{I(j)}^K a_{I(j)}^i \overline{b}_{I(j)}^i \langle Q_z e^i, e^i \rangle \text{ (a.s.c)}$$
and, since $Q_{z,n}[t,a]$ is disspative for each n, we have
$$\langle Q_z[t,a]e^i, e^i \rangle = \langle Q_{z,n}[t,a]e^i, e^i \rangle + \langle [Q_z[t,a] - Q_{z,n}[t,a]]e^i, e^i \rangle \leq \langle [Q_z[t,a] - Q_{z,n}[t,a]]e^i, e^i \rangle$$
Letting $n \to \infty$, we get $\langle Q_z[t,a]e^i, e^i \rangle \leq 0$, so that $\langle \mathbf{Q}_z[t,a]\Phi, \Phi \rangle \leq 0$. The same argument applies to $\mathbf{Q}_z^*[t,a]$. Since $\mathbf{Q}_z[t,a]$ is dissipative and densely defined, it has a (bounded) dissipative closure on $\mathcal{FD}_\otimes$.

It should be noted that the theorem is still true if we allow the approximating sums for condition (3.6) to diverge but at an order less than $\mu_n^{-1+\delta}$, $0 < \delta < 1$, that is, $\sup_t \left| \int_a^t \left( \|\mathbf{H}_z(t)E^i\|^2 - |\langle \mathbf{H}_z(t)E^i, E^i \rangle|^2 \right) ds \right| = \infty$, with
$$\left| \sum_{k,}^n (\Delta t_k)^2 \left( \|H_z(s_k)e^i\|^2 - |\langle H_z(s_k)e^i, e^i \rangle|^2 \right) \right| \leq M\mu_n^\delta.$$

We also note that:

$$\|\mathbf{Q}_z[t,a]\Phi\|_\otimes^2 = \sum_i^J \sum_{I(j)}^K |a_{I(j)}^i|^2 |\langle Q_z e^i, e^i \rangle|^2 \text{ (a.s.c)}, \tag{3.9}$$

in either of the above cases. This representation makes it easy to prove the next theorem.

**Theorem 3.4**

1. $\mathbf{Q}_z[t,s] + \mathbf{Q}_z[s,a] = \mathbf{Q}_z[t,a]$ (a.s.c),
2. $s\text{-}\lim_{h \to 0} \dfrac{\mathbf{Q}_z[t+h,a] - \mathbf{Q}_z[t,a]}{h} = s\text{-}\lim_{h \to 0} \dfrac{\mathbf{Q}_z[t+h,t]}{h} = \mathbf{H}_z(t)$ (a.s.c),
3. $s\text{-}\lim_{h \to 0} \mathbf{Q}_z[t+h,t] = 0$ (a.s.c),

4. $s\text{-}\lim_{h \to 0} \exp\{\tau \mathbf{Q}_z[t+h,t]\} = I_\otimes$ (a.s.c), $\tau \geq 0$.

**Proof:** In each case, it suffices to prove the result for $\Phi \in D_0$. To prove 1., use

$$\left\|[\mathbf{Q}_z[t,s] + \mathbf{Q}_z[s,a]]\Phi\right\|_\otimes^2 = \sum_i^J \sum_{I(j)}^K |a_{I(j)}^i|^2 \left|\langle[Q_z[t,s] + Q_z[s,a]]e^i, e^i\rangle\right|^2 \text{ (a.s.c)}$$

$$= \sum_i^J \sum_{I(j)}^K |a_{I(j)}^i|^2 \left|\langle Q_z[t,a]e^i, e^i\rangle\right|^2 = \left\|\mathbf{Q}_z[t,a]\Phi\right\|_\otimes^2 \text{ (a.s.c)}.$$

To prove 2., use 1 to get $\mathbf{Q}_z[t+h,a] - \mathbf{Q}_z[t,a] = \mathbf{Q}_z[t+h,t]$ (a.s.), so that

$$\lim_{h \to 0} \left\|\frac{\mathbf{Q}_z[t+h,t]}{h}\Phi\right\|_\otimes^2 = \sum_i^J \sum_{I(j)}^K |a_{I(j)}^i|^2 \lim_{h \to 0}\left|\left\langle\frac{Q_z[t+h,t]}{h}e^i, e^i\right\rangle\right|^2 = \left\|\mathbf{H}_z(t)\Phi\right\|_\otimes^2 \text{ (a.s.c.)}.$$

The proof of 3., follows from 2., and the proof of 4. follows from 3.

**Theorem 3.5** *Suppose that* $\lim_{z \to \infty}\langle Q_z[t,a]\phi,\psi\rangle = \langle Q[t,a]\phi,\psi\rangle$ *exists for $\phi$ in a dense set $\forall \psi \in \mathcal{H}$ (weak convergence). Then:*

1. $Q[t,a]$ generates a strongly continuous contraction semigroup on $\mathcal{H}$,

2. $\lim_{z \to \infty} \mathbf{Q}_z[t,a]\Phi = \mathbf{Q}[t,a]\Phi$ for $\Phi \in D_0$ and $\mathbf{Q}[t,a]$ is the generator of a strongly continuous contraction semigroup on $\mathcal{FD}_\otimes$,

3. $\mathbf{Q}[t,s] + \mathbf{Q}[s,a] = \mathbf{Q}[t,a]$ (a.s.c.),
4. $\lim_{h \to 0}\frac{\mathbf{Q}[t+h,a] - \mathbf{Q}[t,a]}{h}\Phi = \lim_{h \to 0}\frac{\mathbf{Q}[t+h,t]}{h}\Phi = \mathbf{H}(t)\Phi$ (a.s.c.),
5. $\lim_{h \to 0}\mathbf{Q}[t+h,t]\Phi = 0$ (a.s.c.), and
6. $\lim_{h \to 0}\exp\{\tau\mathbf{Q}[t+h,t]\}\Phi = \Phi$ (a.s.c.), $\tau \geq 0$.

**Proof:** The proofs are easy. For 1., first note that $Q[t,a]$ is closable and use
$\langle Q[t,a]\phi,\phi\rangle = \langle Q_z[t,a]\phi,\phi\rangle + \langle[Q[t,a] - Q_z[t,a]]\phi,\phi\rangle \leq \langle[Q[t,a] - Q_z[t,a]]\phi,\phi\rangle$ and let

$z \to \infty$. Then do likewise for $\langle \phi, Q^*[t,a]\phi \rangle$ to get that $Q[t,a]$ is maximal dissipative. To prove 2., use (3.9) in the form

$$\left\| [\mathbf{Q}_z[t,a] - \mathbf{Q}_{z'}[t,a]]\Phi \right\|_\otimes^2 = \sum_i^J \sum_{I(j)}^K \left| a_{I(j)}^i \right|^2 \left| \langle [Q_z[t,a] - Q_{z'}[t,a]]e^i, e^i \rangle \right|^2, \text{ (a.s.c.)}.$$

This proves that $\mathbf{Q}_z[t,a] \xrightarrow{s} \mathbf{Q}[t,a]$. Since $\mathbf{Q}[t,a]$ is densely defined, it is closable. The same method as above shows that it is maximal dissipative. Proofs of the other results follow the methods of the previous theorem.

Since $\mathbf{Q}[t,a]$ and $\mathbf{Q}_z[t,a]$ generate contraction semigroups, set $\mathbf{U}[t,a] = \exp\{\mathbf{Q}[t,a]\}$, $\mathbf{U}_z[t,a] = \exp\{\mathbf{Q}_z[t,a]\}$, for $t \in E$. They are evolution operators and the following theorem is a slight modification of a result due to Hille and Phillips[74], known as the second exponential formula.

**Theorem 3.6** *If* $\mathbf{Q}'[t,a] = w\mathbf{Q}[t,a]$ *is the generator of a strongly continuous contraction semigroup, and* $\mathbf{U}^w[t,a] = \exp\{w\mathbf{Q}[t,a]\}$, *then, for each n and* $\Phi \in D\big[(\mathbf{Q}[t,a])^{n+1}\big]$, *we have* (*where* w *is a parameter*)

$$\mathbf{U}^w[t,a]\Phi = \left\{ I_\otimes + \sum_{k=1}^n \frac{(w\mathbf{Q}[t,a])^n}{n!} + \frac{1}{n!} \int_0^w (w-\xi)^n \mathbf{Q}[t,a]^{n+1} \mathbf{U}^\xi[t,a]d\xi \right\} \Phi. \quad (3.10)$$

**Proof:** The proof is easy. Start with $\left[ \mathbf{U}_z^w[t,a]\Phi - I_\otimes \right]\Phi = \int_0^w \mathbf{Q}_z[t,a]\mathbf{U}_z^\xi[t,a]d\xi\Phi$ and use integration by parts to get that

$$\left[ \mathbf{U}_z^w[t,a]\Phi - I_\otimes \right]\Phi = w\mathbf{Q}_z[t,a]\Phi + \int_0^w (w-\xi)[\mathbf{Q}_z[t,a]]^2 \mathbf{U}_z^\xi[t,a]d\xi\Phi.$$

It is clear how to get the n-th term. Finally, let $z \to \infty$ to get (3.10).

**Theorem 3.7** *If* $a < t < b$,

1. $\lim_{z \to \infty} \mathbf{U}_z[t,a]\Phi = \mathbf{U}[t,a]\Phi$, $\Phi \in \mathcal{FD}_\otimes$,
2. $\dfrac{\partial}{\partial t}\mathbf{U}_z[t,a]\Phi = \mathbf{H}_z(t)\mathbf{U}_z[t,a]\Phi = \mathbf{U}_z[t,a]\mathbf{H}_z(t)\Phi$, $\Phi \in \mathcal{FD}_\otimes$, and
3. $\dfrac{\partial}{\partial t}\mathbf{U}[t,a]\Phi = \mathbf{H}(t)\mathbf{U}[t,a]\Phi = \mathbf{U}[t,a]\mathbf{H}(t)\Phi$, $\Phi \in D(\mathbf{Q}[b,a]) \supset D_0$.

**Proof:** To prove 1., use the fact that $\mathbf{H}_z(t)$ and $\mathbf{H}(t)$ commute along with

$$\mathbf{U}[t,a]\Phi - \mathbf{U}_z[t,a]\Phi = \int_0^1 (d/ds)\left(e^{s\mathbf{Q}[t,a]}e^{(1-s)\mathbf{Q}_z[t,a]}\right)\Phi ds$$

$$= \int_0^1 s\left(e^{s\mathbf{Q}[t,a]}e^{(1-s)\mathbf{Q}_z[t,a]}\right)(\mathbf{Q}[t,a] - \mathbf{Q}_z[t,a])\Phi ds, \text{ so that}$$

$$\|\mathbf{U}[t,a]\Phi - \mathbf{U}_z[t,a]\Phi\| \leq \|\mathbf{Q}[t,a]\Phi - \mathbf{Q}_z[t,a]\Phi\|.$$

To prove 2., use

$$\mathbf{U}_z[t+h,a] - \mathbf{U}_z[t,a] = \mathbf{U}_z[t,a](\mathbf{U}_z[t+h,t] - \mathbf{I}) = (\mathbf{U}_z[t+h,t] - \mathbf{I})\mathbf{U}_z[t,a], \text{ so that,}$$

$$\frac{(\mathbf{U}_z[t+h,a] - \mathbf{U}_z[t,a])}{h} = \mathbf{U}_z[t,a]\frac{(\mathbf{U}_z[t+h,t] - \mathbf{I})}{h}.$$

Now set $\Phi_z^t = \mathbf{U}_z[t,a]\Phi$ and use (3.10) with n = 1 and w = 1 to get:

$$\mathbf{U}_z[t+h,t]\Phi_z^t = \left\{\mathbf{I}_\otimes + \mathbf{Q}_z[t+h,t] + \int_0^1 (1-\xi)\ \mathbf{U}_z^\xi[t+h,t]\mathbf{Q}_z[t+h,t]^2 d\xi\right\}\Phi_z^t, \text{ so that}$$

$$\frac{(\mathbf{U}_z[t+h,t] - \mathbf{I})}{h}\Phi_z^t - \mathbf{H}_z(t)\Phi_z^t = \frac{\mathbf{Q}_z[t+h,t]}{h}\Phi_z^t - \mathbf{H}_z(t)\Phi_z^t$$

$$+ \int_0^1 (1-\xi)\ \mathbf{U}_z^\xi[t+h,t]\frac{\mathbf{Q}_z[t+h,t]^2}{h}\ \Phi_z^t d\xi.$$

It follows that

$$\left\|\frac{(\mathbf{U}_z[t+h,t] - \mathbf{I})}{h}\Phi_z^t - \mathbf{H}_z(t)\Phi_z^t\right\|_\otimes \leq \left\|\frac{\mathbf{Q}_z[t+h,t]}{h}\Phi_z^t - \mathbf{H}_z(t)\Phi_z^t\right\|_\otimes + \frac{1}{2}\left\|\frac{\mathbf{Q}_z[t+h,t]^2}{h}\ \Phi_z^t\right\|_\otimes.$$

The result now follows from Theorem (3.4)-2 and 3.

To prove 3., note that $\mathbf{H}_z(t)\Phi = \mathbf{H}(t)\{z\mathbf{R}(z,\mathbf{H}(t))\}\Phi = \{z\mathbf{R}(z,\mathbf{H}(t))\}\mathbf{H}(t)\Phi$, so that $\{z\mathbf{R}(z,\mathbf{H}(t))\}$ commutes with $\mathbf{U}[t,a]$ and $\mathbf{H}(t)$. Now show that

$$\|\mathbf{H}_z(t)\mathbf{U}_z[t,a]\Phi - \mathbf{H}_{z'}(t)\mathbf{U}_{z'}[t,a]\Phi\| \leq \|[\mathbf{U}_z[t,a]\Phi - \mathbf{U}_{z'}[t,a]]\mathbf{H}(t)\Phi\|$$
$$+ \|[z\mathbf{R}(z,\mathbf{H}(t))\Phi - z'\mathbf{R}(z',\mathbf{H}(t))]\mathbf{H}(t)\Phi\| \to 0, \ z,z' \to \infty,$$

so that, for $\Phi \in D(\mathbf{Q}[b,a])$, $\mathbf{H}_z(t)\mathbf{U}_z[t,a]\Phi \to \mathbf{H}(t)\mathbf{U}[t,a]\Phi = \dfrac{\partial}{\partial t}\mathbf{U}[t,a]\Phi$.

The previous theorems form the core of our approach to the Feynman operator calculus. Our theory applies to both hyperbolic and parabolic equations. In the conventional approach, these two cases require different methods (see Pazy[71]). It is not hard to show that the requirements imposed in these cases are stronger than (our condition of) weak integral. This will be discussed in a later paper devoted to the general problem on Banach spaces.

## 4. Perturbation Theory

**Definition 4.1** The evolution operator $\mathbf{U}^w[t,a] = \exp\{w\mathbf{Q}[t,a]\}$, is said to be <u>asymptotic in the sense of Poincaré</u>, if for each n and each $\Phi_a \in D\big[(\mathbf{Q}[t,a])^{n+1}\big]$, we have

$$\lim_{w \to 0} w^{-(n+1)}\left\{\mathbf{U}^w[t,a] - \sum_{k=1}^{n}\frac{(w\mathbf{Q}[t,a])^k}{k!}\right\}\Phi_a = \frac{\mathbf{Q}[t,a]^{n+1}}{(n+1)!}\Phi_a. \tag{4.1}$$

This is the operator version of an asymptotic expansion in the classical sense, but here $\mathbf{Q}[t,a]$ is an unbounded operator.

As noted earlier, Dyson[16] analyzed the (renormalized) perturbation expansion for quantum electrodynamics and suggested that it actually diverges. He concluded that we could, at best, hope that the series is asymptotic. His arguments were based on (not completely convincing) physical considerations, but no precise formulation of the problem was possible at that time. However, the calculations of Hurst[75], Thirring[76], Peterman[77], and Jaffe[78] for specific models all support Dyson's contention that the renormalized perturbation series diverges. In his recent book[91] (pg. 13-16), Dyson's views on the perturbation series and renormalization are reiterated: " … in spite of all the successes of the new physics, the two questions that defeated me in 1951 remain unsolved." Here, he is referring to the question of mathematical consistency for the whole renormalization program, and our ability to (reliably) calculate nuclear processes in quantum chromodynamics. (For other details and references to additional works, see Schweber[6,80], Wightman[84] and Zinn-Justin[79].)

The general construction of a physically simple and mathematically satisfactory formulation of quantum electrodynamics is still an open problem. The next theorem establishes Dyson's (second) conjecture under conditions that would apply to any (future) theory that does not require a radical departure from the present foundations of quantum theory (unitary solution operators). It also applies to the renormalized expansions in some areas of condensed matter physics where the solution operators are contraction semigroups.

**Theorem 4.2** *Suppose the conditions for Theorem 3.5 are satisfied. Then:*

1. $\mathbf{U}^w[t,a] = \exp\{w\mathbf{Q}[t,a]\}$ *is asymptotic in the sense of Poincaré.*

2. *For each n and each* $\Phi_a \in D[(\mathbf{Q}[t,a])^{n+1}]$, *we have*

$$\Phi(t) = \Phi_a + \sum_{k=1}^{n} w^k \int_a^t ds_1 \int_a^{s_1} ds_2 \cdots \int_a^{s_{k-1}} ds_k \mathbf{H}(s_1)\mathbf{H}(s_2)\cdots\mathbf{H}(s_k)\Phi_a \qquad (4.2)$$
$$+ \int_0^w (w-\xi)^n d\xi \int_a^t ds_1 \int_a^{s_1} ds_2 \cdots \int_a^{s_n} ds_{n+1} \mathbf{H}(s_1)\mathbf{H}(s_2)\cdots\mathbf{H}(s_{n+1})\mathbf{U}^\xi[s_{n+1},a]\Phi_a,$$

where $\Phi(t) = \mathbf{U}^w[t,a]\Phi_a$.

**Proof:** From (3.10), we have

$$\mathbf{U}^w[t,a]\Phi = \left\{ \sum_{k=0}^{n} \frac{(w\mathbf{Q}[t,a])^n}{n!} + \frac{1}{n!}\int_0^w (w-\xi)^n \mathbf{Q}[t,a]^{n+1} \mathbf{U}^\xi[t,a] d\xi \right\}\Phi,$$

so that

$$w^{-(n+1)}\left\{ \mathbf{U}^w[t,a]\Phi_a - \sum_{k=0}^{n} \frac{(w\mathbf{Q}[t,a])^k}{k!}\Phi_a \right\} = +\frac{(n+1)}{(n+1)!} w^{-(n+1)} \int_0^w (w-\xi)^n d\xi \mathbf{U}^\xi[t,a]\mathbf{Q}[t,a]^{n+1}\Phi_a.$$

Replace the right hand side by

$$I = \frac{(n+1)}{(n+1)!} w^{-(n+1)} \int_0^w (w-\xi)^n d\xi \left\{ \mathbf{U}_z^\xi[t,a] + \left[\mathbf{U}^\xi[t,a] - \mathbf{U}_z^\xi[t,a]\right]\right\} \mathbf{Q}[t,a]^{n+1}\Phi_a.$$

Now, expand the term $\mathbf{U}_z^\xi[t,a]$ in a two-term Taylor series about zero to get

$$\mathbf{U}_z^\xi[t,a] = I_\otimes + \xi \mathbf{Q}_z[t,a] + R_z^\xi.$$

Put the above in *I*, compute the elementary integrals showing that only the $I_\otimes$ term gives a nonzero value (of $1/(n+1)$) when $w \to 0$. Then let $z \to \infty$ to get

$$\lim_{w \to 0}(n+1)w^{-(n+1)}\int_0^w d\xi(w-\xi)^n \mathbf{U}^\xi[t,a]\mathbf{Q}[t,a]^{n+1}\Phi_a = \mathbf{Q}[t,a]^{n+1}\Phi_a.$$

This proves that $\mathbf{U}[t,a] = \exp\{\mathbf{Q}[t,a]\}$ is asymptotic in the sense of Poincaré. To prove (4.2), let $\Phi_a \in D\left[(\mathbf{Q}[t,a])^{n+1}\right]$ for each $k \leq n+1$, and use the fact that (Dollard and Friedman[81])

$$(\mathbf{Q}_z[t,a])^k \Phi_a = \left(\int_a^t \mathbf{H}_z(s)ds\right)^k \Phi_a = (k!)\int_a^t ds_1 \int_a^{s_1} ds_2 \cdots \int_a^{s_{k-1}} ds_n \mathbf{H}_z(s_1)\mathbf{H}_z(s_2)\cdots\mathbf{H}_z(s_k)\Phi_a. \quad (4.3)$$

Letting $z \to \infty$ gives the result.

Our conditions are very weak. For example, the recent work of Tang and Li[82] required that $\|H(t)\|$ be Lebesgue integrable.

There are well known special cases in which the perturbation series may actually converge to the solution. This can happen, for example, if the generator is bounded or if it is analytic in some sector. More generally, when the generator is of the form $\mathbf{H}(t) = \mathbf{H}_0(t) + \mathbf{H}_i(t)$, where $\mathbf{H}_0(t)$ is analytic and $\mathbf{H}_i(t)$ is some reasonable perturbation, which need not be bounded, there are conditions that allow the interaction representation to have a convergent Dyson expansion. These results can be formulated and proven in our formalism. However, the proofs are essentially the same as in the standard case so we will present them in a later paper devoted to the operator calculus on Banach spaces. The recent book by Engel and Nagel[83] provides some new results in this general area.

There are also cases where the (renormalized) series may diverge, but still respond to some summability method. This phenomenon is well known in classical analysis. In field theory, things can be much more complicated. A good discussion, with references, can be found in the review by Wightman[84] and the book by Glimm and Jaffe[34].

## 5. Sum Over Paths

In this section we first review and make a distinction between what is actually known and what we think we know about the foundations for our physical view of the micro-world. The objective is to provide the background for a number of physically motivated postulates that will be used to develop a theory of measurement for the micro-world (sufficient for our purposes). This will allow us to relate the theory of Sections 3 and 4 to Feynman's sum over paths approach and prove Dyson's second conjecture. This section differs from the previous ones in that we shift the orientation and perspective from that of mathematical physics to that of theoretical physics.

In spite of the enormous successes of the physical sciences in the past century, our information and understanding about the micro-world is still rather meager. In the macro-world we are quite comfortable with the view that physical systems evolve continuously in time and our results justify this view. Indeed, the success of continuum physics is the basis for a large part of our technical advances in the twentieth century. On the other hand, the same view is also held at the micro-level and, in this case, our position is not very secure. The ability to measure physical events continuously in time at the micro-level must be considered a belief which, although convenient, has no place in science as an a priori constraint.

In order to establish perspective, let us consider this belief within the context of a satisfactory, and well-justified theory, Brownian motion. This theory lies at the interface between the macro- and the micro-worlds. Some presentations of this theory (the careful ones) make a distinction between the mathematical and the physical foundations of Brownian motion and that distinction is important for our discussion.

When Einstein[85] began his investigation of the physical issues associated with this phenomenon, he was forced to assume that physical information about the state of a Brownian particle (position, velocity, etc) can only be known in time intervals that are large compared with the mean time between molecular collisions. (It is known that, under normal physical conditions, a Brownian particle receives about $10^{21}$ collisions per second.) Wiener took the mathematical step and assumed that this mean time (between collisions) could be made zero, thus providing a mathematical Brownian particle. This corresponds physically to the assumption that the ratio of the mass of the particle to the friction of the fluid is zero in the limit (see Wiener et al[86]).

From the physical point of view, use of Wiener's idealization of the Einstein model was not satisfactory since it led to problems of unbounded path length and nondifferentiability at all points. The first problem is physically impossible while the second is physically unreasonable. Of course, the idealization has turned out to be quite satisfactory in areas where the information required need not be detailed, such as large parts of electrical engineering, chemistry, and the biological sciences. Ornstein and Uhlenbeck[87] later constructed a model that, gives the Einstein view asymptotically, but in small-time regions, is equivalent to the assumption that the particle travels a linear path between collisions. This model provides finite path length and differentiability. (The theory was later idealized by Doob[88].) What we do know is that the very nature of the liquid state implies collective behavior among the molecules. *This means that we do not know what path the particle travels in between collisions*. However, since the tools and methods of analysis require some form of continuity, some such (in between observation) assumptions must be made. It is clear that the need for these assumptions is imposed by the available mathematical structures within which we must represent physical reality as a model.

Theoretical science concerns itself with the construction of mathematical representations of certain restricted portions of physical reality. Various trends and philosophies that are prevalent at the time temper these constructs. A consistent theme has been the quest for simplicity. This requirement is born out of the natural need to restrict models to the minimum number of variables, relationships, constraints, etc, which give a satisfactory account of known experimental results and possibly allow the prediction of heretofore unknown consequences. One important outcome of this approach has been to implicitly eliminate all reference to the background within which physical systems evolve. In the micro-world, such an action cannot be justified without prior investigation. We propose to replace the use of mathematical coordinate systems by "physical coordinate systems" in order to (partially) remedy this problem.

We denote a physical coordinate system at time $t$ by $\mathbf{R}_p^3(t)$. This coordinate system is attached to an observer (including measuring devices) and is envisioned as $\mathbf{R}^3$ plus any background effects, either local or distant, which affect the observer's ability to obtain precise (ideal) experimental information about physical reality. This in turn affects our observer's ability to construct precise (ideal) representations and make precise predictions about physical reality (in the micro-world).

More specifically, consider the evolution of some micro-system on the interval $E = [a,b]$. Physically this evolution manifests itself as a curve on $\mathbf{X}$, where

$$\prod_{t \in E} \mathbf{R}_p^3(t) = \mathbf{X}.$$

Thus, true physical events occur on $\mathbf{X}$ where actual experimental information is modified by fluctuations in $\mathbf{R}_p^3(t)$, and by the interaction of the micro-system with the measuring equipment. Based on the success of our models, we know that such small changes are in the noise region, and they have no effect on our predictions for macro-systems. However, there is no a priori reason to believe that the effects will be small on micro-systems.

In terms of our theoretical representations, we are forced to model the evolution of physical systems in terms of wave functions, amplitudes, and/or operator-valued distributions, etc. There are thus two spaces, the physical space of evolution for the micro-system and the observer's space of obtainable information concerning this evolution. The lack of distinction between these two spaces seems to be the cause for some of the confusion and lack of physical clarity. For example, it may be perfectly correct to assume that a particle travels a continuous path on $\mathbf{X}$. However, the assumption that the observer's space of obtainable information includes infinitesimal spacetime knowledge of this path is completely unfounded. This leads to our first postulate:

**Postulate 1.** *Physical reality is a continuous process in time.*

We thus take this view, fully recognizing that experiment does not provide continuous information about physical reality, and that there is no reason to believe that our mathematical representations contain precise information about the continuous spacetime behavior of physical processes at this level.

Since the advent of the special theory of relativity, there is much discussion about events, which generally means a point in $\mathbf{R}^4$ with the Minkowski metric. In terms of real physics, this is a fiction which is frequently useful for reasons of presentation but so widely used that, to avoid confusion, it is appropriate to define what we mean by a *physical event*.

**Definition 5.1** *A physical event is a set of physical changes in a given system that can be verified directly by experiment or indirectly via subsequent changes, where conclusions are based on an a priori agreed-upon model of the physical process.*

This definition corresponds more closely to what is meant by physical events. It explicitly recognizes the evolution of scientific inference and the need for general agreement about what is being observed (based on specific models).

Before continuing, it will be helpful to have a particular physical picture in mind that makes the above discussion explicit. For this purpose, we take this picture to be a photograph showing the track left by a π-meson in a bubble chamber (and take seriously the amount of information available). In particular, we assume that the following reaction occurs:

$$\pi^+ \to \mu^+ + \nu.$$

We further assume that the orientation of our photograph is such that the π-meson enters on the left at time $t=0$ and the tracks left by the µ-meson disappear on the right at time $t=T$, where $T$ is of the order of $10^{-3}$sec, the time exposure for photographic film. Although the neutrino does not appear in the photograph, we also include a track for it. In Figure I we present a simplified picture of this photograph.

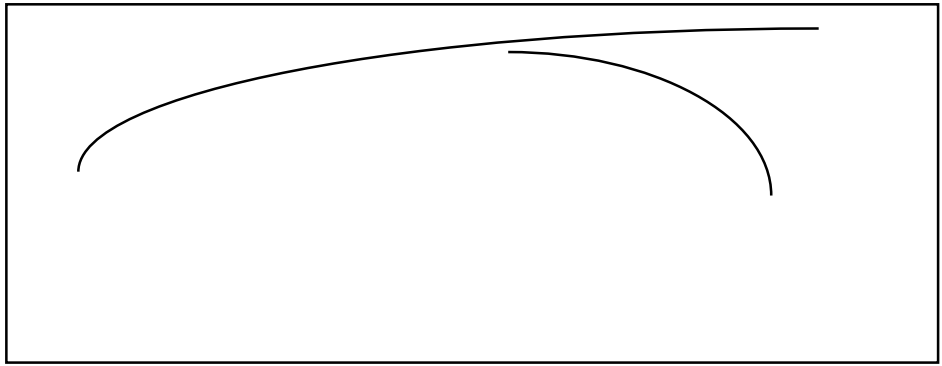

**Figure I**

We have drawn the photograph as if we continuously see the particles in the picture. However, experiment only provides us individual bubbles, which do not necessarily overlap, from which we must extract physical information. A more accurate (though still not realistic) depiction is given in Figure II.

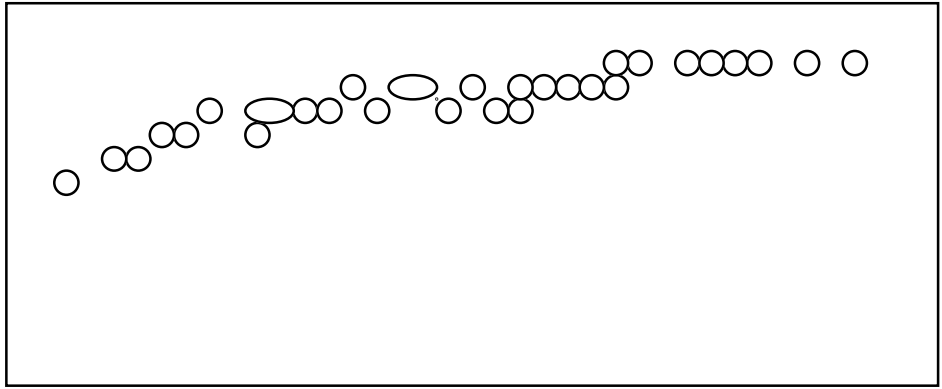

**Figure II**

Let us assume that we have magnified a portion of our photograph to the extent that we may distinguish the individual bubbles created by the π-meson as it passes through the chamber. In Figure III, we present a simplified model of adjacent bubbles.

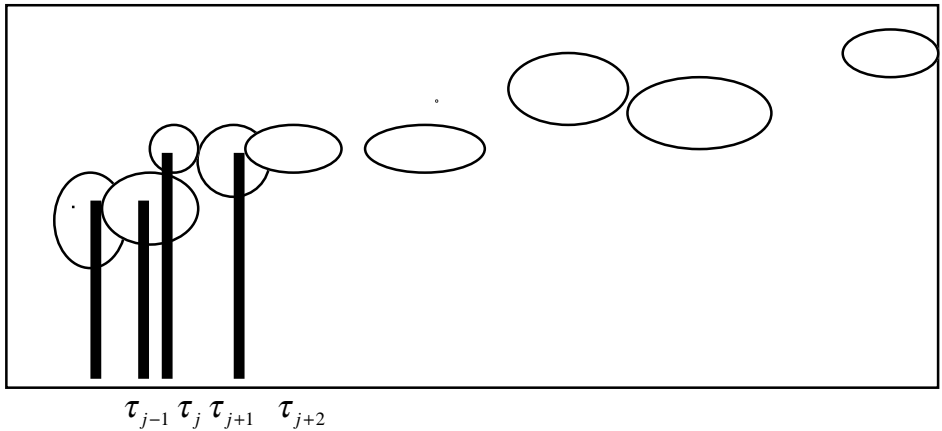

$\tau_{j-1}\ \tau_j\ \tau_{j+1}\quad \tau_{j+2}$

**Figure III**

**Postulate 2.** *We assume that the center of each bubble represents the average knowable effect of the particle in a symmetric time interval about the center.*

By average knowable effect, we mean the average of the physical observables. In Fig III, we consider the existence of a bubble at time $\tau_j$ to be caused by the average of the physical observables over the time interval $[t_{j-1}, t_j]$, where $t_{j-1} = (1/2)[\tau_{j-1} + \tau_j]$ and $t_j = (1/2)[\tau_j + \tau_{j+1}]$. This postulate requires some justification. In general, the resolution of film and the relaxation time for distinct bubbles in the chamber vapor are limited. This means that if the π-meson creates two bubbles that are closely spaced in time, the bubbles may coalesce and appear as one. If this does not occur, it is still possible that the film will record the event as one bubble because of its inability to resolve events is such small time intervals.

Let us now recognize that we are dealing with one photograph so that, in order to obtain all available information, we must analyze a large number of photographs of the same reaction obtained under similar conditions (pre-prepared states). It is clear that the number of bubbles and the time placement of the bubbles will vary (independently of each other) from photograph to photograph. Let $\lambda^{-1}$ denote the average time for the appearance of a bubble in the film.

**Postulate 3.** *We assume that the number of bubbles in any film is a random variable.*

**Postulate 4.** *We assume that, given that n bubbles have appeared on a film, the time positions of the centers of the bubbles are uniformly distributed.*

**Postulate 5.** *We assume that N(t), the number of bubbles up to time t in a given film, is a Poisson-distributed random variable with parameter $\lambda$.*

To motivate Postulate 5, recall that $\tau_j$ is the time center of the j-th bubble and $\lambda^{-1}$ is the average (experimentally determined) time between bubbles. The following results can be found in Ross[89].

**Theorem 5.1** *The random variables $\Delta\tau_j = \tau_j - \tau_{j-1}$ ($\tau_0 = 0$) are independent identically distributed random variables of exponential type with mean $\lambda^{-1}$, for $1 \leq j \leq n$.*

The arrival times $\tau_1, \tau_2, \cdots, \tau_n$ are not independent, but their density function can be computed from

$$\mathrm{Prob}[\tau_1, \tau_2, \cdots, \tau_n] = \mathrm{Prob}[\tau_1]\mathrm{Prob}[\tau_2 \mid \tau_1] \cdots \mathrm{Prob}[\tau_n \mid \tau_1, \tau_2, \cdots, \tau_{n-1}]. \qquad (5.1a)$$

We now use Theorem 5.1 to conclude that, for $k \geq 1$,

$$\mathrm{Prob}[\tau_k \mid \tau_1, \tau_2, \cdots, \tau_{k-1}] = \mathrm{Prob}[\tau_k \mid \tau_{k-1}]. \qquad (5.1b)$$

We don't know this conditional probability however, the natural assumption is that given n bubbles appear, they are equally (uniformly) distributed on the interval. We can now construct what we call the experimental evolution operator. Assume that the conditions for Theorem 3.5 are satisfied and that the family $\{\tau_1, \tau_2, \cdots, \tau_n\}$ represents the time positions of the centers of n bubbles in our film of Fig III. Set $a = 0$ and define $\mathbf{Q}_E[\tau_1, \tau_2, \cdots, \tau_n]$ by

$$\mathbf{Q}_E[\tau_1, \tau_2, \cdots, \tau_n] = \sum_{j=1}^{n} \int_{t_{j-1}}^{t_j} E[\tau_j, s]\mathbf{H}(s)ds. \qquad (5.2a)$$

Here, $t_0 = \tau_0 = 0$, $t_j = (1/2)[\tau_j + \tau_{j+1}]$ (for $1 \leq j \leq n$), and $E[\tau_j, s]$ is the exchange operator defined in Section 2. The effect of our exchange operator $E[\tau_j, s]$ is to concentrate all information contained in $[t_{j-1}, t_j]$ at $\tau_j$. This is how we implement our postulate that the known physical event of the bubble at time $\tau_j$ is due to an average of physical effects over $[t_{j-1}, t_j]$ with information concentrated at $\tau_j$. We can rewrite $\mathbf{Q}_E[\tau_1, \tau_2, \cdots, \tau_n]$ as

$$\mathbf{Q}_E[\tau_1, \tau_2, \cdots, \tau_n] = \sum_{j=1}^{n} \Delta t_j \left[ \frac{1}{\Delta t_j} \int_{t_{j-1}}^{t_j} E[\tau_j, s] \mathbf{H}(s) ds \right]. \quad (5.2b)$$

Thus, we indeed have an average as required by Postulate 2. The evolution operator is given by

$$U[\tau_1, \tau_2, \cdots, \tau_n] = \exp\left\{ \sum_{j=1}^{n} \Delta t_j \left[ \frac{1}{\Delta t_j} \int_{t_{j-1}}^{t_j} E[\tau_j, s] \mathbf{H}(s) ds \right] \right\}. \quad (5.3a)$$

For $\Phi \in \mathcal{FD}_\otimes$, we define the function $\mathbf{U}[N(t), 0]\Phi$ by:

$$\mathbf{U}[N(t), 0]\Phi = U[\tau_1, \tau_2, \cdots, \tau_{N(t)}]\Phi. \quad (5.3b)$$

The function $\mathbf{U}[N(t), 0]\Phi$ is a $\mathcal{FD}_\otimes$-valued random variable, which represents the distribution of the number of bubbles that may appear on our film up to time $t$. In order to relate $\mathbf{U}[N(t), 0]\Phi$ to actual experimental results, we must compute its expected value. Using Postulates 3, 4, and 5, we have

$$\overline{\mathbf{U}}_\lambda[t, 0]\Phi = \mathcal{E}[\mathbf{U}[N(t), 0]\Phi] = \sum_{n=0}^{\infty} \mathcal{E}\{\mathbf{U}[N(t), 0]\Phi | N(t) = n\} \text{Prob}[N(t) = n], \quad (5.4a)$$

$$\mathcal{E}\{\mathbf{U}[N(t), 0]\Phi | N(t) = n\} = \int_0^t \frac{d\tau_1}{t} \int_{\tau_1}^t \frac{d\tau_2}{t - \tau_1} \cdots \int_{\tau_{n-1}}^t \frac{d\tau_n}{t - \tau_{n-1}} U[\tau_n, \cdots, \tau_1]\Phi = \overline{\mathbf{U}}_n[t, 0]\Phi, \quad (5.5a)$$

and

$$\text{Prob}[N(t) = n] = \frac{(\lambda t)^n}{n!} \exp\{-\lambda t\}. \tag{5.6}$$

The integral in (5.4a) acts to distribute uniformly the time positions $\tau_j$ over the successive intervals $[t, \tau_{j-1}]$, $1 \leq j \leq n$, given that $\tau_{j-1}$ has been determined. This is a natural result given our lack of knowledge.

The integral (5.4a) is of theoretical value but is not easy to compute. Since we are only interested in what happens when $\lambda \to \infty$, and as the mean number of bubbles in the film at time $t$ is $\lambda t$, we can take $\tau_j = (jt/n)$, $1 \leq j \leq n$, ($\Delta t_j = t/n$ for each n). We can now replace $\overline{\mathbf{U}}_n[t,0]\Phi$ by $\mathbf{U}_n[t,0]\Phi$, and with this understanding, we continue to use $\tau_j$, so that

$$\mathbf{U}_n[t,0]\Phi = \exp\left\{\sum_{j=1}^{n} \int_{t_{j-1}}^{t_j} E[\tau_j, s]\mathbf{H}(s)ds\right\}\Phi. \tag{5.5b}$$

We define our experimental evolution operator $\mathbf{U}_\lambda[t,0]\Phi$ by

$$\mathbf{U}_\lambda[t,0]\Phi = \sum_{n=0}^{\infty} \frac{(\lambda t)^n}{n!} \exp\{-\lambda t\}\mathbf{U}_n[t,0]\Phi. \tag{5.4b}$$

We now have the following result, which is a consequence of the fact that Borel summability is regular.

**Theorem 5.4** *Assume that the conditions for Theorem 3.5 are satisfied. Then*

$$\lim_{\lambda \to \infty} \overline{\mathbf{U}}_\lambda[t,0]\Phi = \lim_{\lambda \to \infty} \mathbf{U}_\lambda[t,0]\Phi = \mathbf{U}[t,0]\Phi. \tag{5.7}$$

Since $\lambda \to \infty \Rightarrow \lambda^{-1} \to 0$, this means that the average time between bubbles is zero (in the limit) so that we get a continuous path. It should be observed that this continuous path arises from averaging the sum over an infinite number of (discrete) paths. The first term in (5.4b) corresponds to the path of a π-meson that created no bubbles (i.e., the photograph is blank). This event has probability exp{−λt} (which approaches zero as $\lambda \to \infty$). The n-th term corresponds to the path of a π-meson that created n bubbles, (with probability $[(\lambda t)^n / n!]\exp\{-\lambda t\}$) etc. Before deriving a physical relationship, let $P[t;s,\lambda] = 0$ if $s \leq 0$ and, for $0 < s < \infty$, define it as:

$$P[t;s,\lambda] = e^{-\lambda t} \sum_{k=0}^{\lceil \lambda s \rceil} \frac{(\lambda t)^k}{k!}, \qquad (5.8)$$

where $n = \lceil \lambda s \rceil$ is the greatest integer $\leq \lambda s$. We can now write $\mathbf{U}[t,0]\Phi$ as

$$\begin{aligned}
\mathbf{U}[t,0]\Phi &= \lim_{\lambda \to \infty} \int_0^\infty d_s P[t;s,\lambda] \mathbf{U}_{\lceil \lambda s \rceil}[s,0]\Phi, \\
\mathbf{U}_{\lceil \lambda s \rceil}[s,0]\Phi &= \exp\left\{\sum_{j=1}^{\lceil \lambda s \rceil} \int_{t_{j-1}}^{t_j} E[\tau_j,u]\mathbf{H}(u)du\right\}\Phi
\end{aligned} \qquad (5.9)$$

Equation (5.9) means that we get both a sum over paths and a probability interpretation for our formalism. This allows us to give a new definition for path integrals.

Suppose the evolution operator $\mathbf{U}[t,0]$ has a kernel, $\mathbf{K}[\mathbf{x}(t), t; \mathbf{x}(0), 0]$, such that

1. $\mathbf{K}[\mathbf{x}(t),\ t;\ \mathbf{x}(s),\ s] = \int_{\mathbf{R}^3} \mathbf{K}[\mathbf{x}(t),\ t;\ \mathbf{x}(s),\ s]\mathbf{K}[\mathbf{x}(s),\ s;\ \mathbf{x}(0),\ 0]d\mathbf{x}(s)$, and
2. $\mathbf{U}[t,0]\Phi = \int_{\mathbf{R}^3} \mathbf{K}[\mathbf{x}(t),\ t;\ \mathbf{x}(0),\ 0]d\mathbf{x}(0)$.

Then, from equation (5.9), we have that

$$\mathbf{U}[t,0]\Phi = \lim_{\lambda \to \infty} \int_0^\infty d_s P[t;s,\lambda]\left\{\prod_{j=1}^{\lceil \lambda s \rceil} \int_{\mathbf{R}^3} \mathbf{K}\big[\mathbf{x}(t_j),\ t_j;\ \mathbf{x}(t_{j-1}),\ t_{j-1}\big] \prod_{j=1}^{\lceil \lambda s \rceil} d\mathbf{x}(t_{j-1})\Phi(0)\right\}.$$

Thus, whenever we can associate a kernel with our evolution operator, the time-ordered version always provides a well-defined path-integral as a sum over paths.  The definition does not (directly) depend on the space of continuous paths and is independent of a theory of measure on infinite dimensional spaces.  Feynman suggested that the operator calculus was more general, in his book with Hibbs[90] (see pg. 355-6).

## 6. The S-Matrix

The objective of this section is to provide a formulation of the S-matrix that will allow us to investigate the sense in which we can believe Dyson's first conjecture.  At the end of his second paper on the relationship between the Feynman and Schwinger-Tomonaga theories, he explored the difference between the divergent Hamiltonian formalism that one must begin with and the finite S-matrix that results from renormalization.  He takes the view that it is a contrast between a real observer and a fictitious (ideal) observer.  The real observer can only determine particle positions with limited accuracy and always gets finite results from his measurements.  Dyson then suggests that "... The ideal observer, however, using non-atomic apparatus whose location in space and time is known with infinite precision, is imagined to be able to disentangle a single field from its interactions with others, and to measure the interaction.  In conformity with the Heisenberg uncertainty principle, it can perhaps be considered a physical consequence of the infinitely precise knowledge of (particle) location allowed to the ideal observer, that the value obtained when he measures (the interaction) is infinite."  He goes on to remark that if his analysis is correct, the problem of divergences is attributable to an idealized concept of measurability.

In order to explore this idea, we work in the interaction representation with obvious notation. Replace the interval $[t, 0]$ by $[T, -T]$, $\mathbf{H}(t)$ by $(-i/\hbar)\mathbf{H}_I(t)$, and our experimental evolution operator $\mathbf{U}_\lambda[T,-T]\Phi$ by the experimental scattering operator $\mathbf{S}_\lambda[T,-T]\Phi$, where

$$\mathbf{S}_\lambda[T,-T]\Phi = \sum_{n=0}^{\infty} \frac{(2\lambda T)^n}{n!} \exp[-2\lambda T]\mathbf{S}_n[T,-T]\Phi, \tag{6.1}$$

$$\mathbf{S}_n[T,-T]\Phi = \exp\left\{(-i/\hbar)\sum_{j=1}^{n}\int_{t_{j-1}}^{t_j} E[\tau_j,s]\mathbf{H}_I(s)ds\right\}\Phi, \tag{6.2}$$

and $\mathbf{H}_I(t) = \int_{\mathbf{R}^3} \mathbf{H}_I(\mathbf{x}(t),t)d\mathbf{x}(t)$ is the interaction energy. We follow Dyson for consistency (see also the discussion), so that $\delta mc^2$ is the mass counter-term designed to cancel the self-energy divergence, and

$$\mathbf{H}_I(\mathbf{x}(t),t) = -ie\mathbf{A}_\mu(\mathbf{x}(t),t)\overline{\psi}(\mathbf{x}(t),t)\gamma_\mu\psi(\mathbf{x}(t),t) - \delta mc^2\overline{\psi}(\mathbf{x}(t),t)\psi(\mathbf{x}(t),t). \tag{6.3}$$

We now give a physical interpretation of our formalism. Rewrite equation (6.1) as

$$\mathbf{S}_\lambda[T,-T]\Phi = \sum_{n=0}^{\infty} \frac{(2\lambda T)^n}{n!} \exp\left\{(-i/\hbar)\sum_{j=1}^{n}\int_{t_{j-1}}^{t_j} \left[E[\tau_j,s]\mathbf{H}_I(s) - i\lambda\hbar\mathbf{I}_\otimes\right]ds\right\}\Phi. \tag{6.4}$$

In this form, it is clear that the term $-i\lambda\hbar\mathbf{I}_\otimes$ has a physical interpretation as the absorption of photon energy of amount $\lambda\hbar$ in each subinterval $[t_j,t_{j-1}]$ (cf. Mott and Massey[92]). When we compute the limit, we get the standard S-matrix (on [*T, -T*]). It follows that we must add an infinite amount of photon energy to the mathematical description of the experimental picture (at each point in time) in order to obtain the standard scattering operator. This is the ultraviolet divergence and shows explicitly that the transition from the experimental to the ideal scattering operator requires that we illuminate the particle throughout its entire path. Thus, it appears that we have, indeed, violated the uncertainty relation. This is further supported if we look at the form of the standard S-matrix:

$$\mathbf{S}[T,-T]\Phi = \exp\left\{(-i/\hbar)\int_{-T}^{T}\mathbf{H}_{I}(s)ds\right\}\Phi, \qquad (6.5)$$

and note that the differential *ds* in the exponent implies perfect infinitesimal time knowledge at each point, strongly suggesting that the energy should be totally undetermined. If violation of the Heisenberg uncertainty relation is the cause for the ultraviolet divergence, then as it is a variance relation, it will not appear in first order (perturbation) but should show up in all higher-order terms. On the other hand, if we eliminate the divergent terms in second order, we would expect our method to prevent them from appearing in any higher order term of the expansion. The fact that this is precisely the case in quantum electrodynamics is a clear verification of Dyson's conjecture.

If we allow T to become infinite, we once again introduce an infinite amount of energy into the mathematical description of the experimental picture, as this is also equivalent to precise time knowledge (at infinity). Of course, this is the well-known infrared divergence and can be eliminated by keeping $T$ finite (see Dahmen et al[93]) or introducing a small mass for the photon (see Feynman[12], pg. 769). If we hold $\lambda$ fixed while letting T become infinite, the experimental S-matrix takes the form:

$$\mathbf{S}_\lambda[\infty,-\infty]\Phi = \exp\left\{(-i/\hbar)\sum_{j=1}^{\infty}\int_{t_{j-1}}^{t_j} E[\tau_j,s]\mathbf{H}_I(s)ds\right\}\Phi,$$

$$\bigcup_{j=1}^{\infty}[t_{j-1},t_j] = (-\infty,\infty), \ \& \ \Delta t_j = \lambda^{-1}.$$

(6.6)

This form is interesting since it shows how a minimal time eliminates the ultraviolet divergence. Of course, this is not unexpected, and has been known at least since Heisenberg[94] introduced his fundamental length as a way around the divergences. This was a prelude to the various lattice approximation methods. The review by Lee[95] is interesting in this regard.

In closing this section, we record our exact expansion for the S-matrix to any finite order. With $\Phi(-\infty) \in D\left[(\mathbf{Q}[\infty,-\infty])^{n+1}\right]$, we have

$$\mathbf{S}[\infty,-\infty]\Phi(-\infty) = \sum_{k=0}^{n}\left(\frac{-i}{\hbar}\right)^k \int_{-\infty}^{\infty}ds_1\int_{-\infty}^{s_1}ds_2\cdots\int_{-\infty}^{s_{k-1}}ds_k\mathbf{H}_I(s_1)\mathbf{H}_I(s_2)\cdots\mathbf{H}_I(s_k)\Phi(-\infty)$$

$$+\left(\frac{-i}{\hbar}\right)^{n+1}\int_0^1(1-\xi)^n d\xi\int_{-\infty}^{\infty}ds_1\int_{-\infty}^{s_1}ds_2\cdots\int_{-\infty}^{s_n}ds_{n+1}\mathbf{H}_I(s_1)\mathbf{H}_I(s_2)\cdots\mathbf{H}_I(s_{n+1})\mathbf{S}^\xi[s_{n+1},-\infty]\Phi(-\infty).$$

(6.7)

It follows that (in a theoretical sense) we can consider the standard S-matrix expansion to be exact, when truncated at any order, by adding the last term of equation (6.7) to give the remainder. This result also means that whenever we can construct an exact nonperturbative solution, it always implies the existence of a perturbative solution valid to any order. However, in general, only in particular cases can we know if the series at some n (without the remainder) approximates the solution.

**Discussion**

In this paper we have shown how to construct a natural representation Hilbert space for Feynman's time-ordered operator calculus. This space allows us to construct the time-ordered integral and evolution operator (propagator) under the weakest known conditions. Using the theory, we have shown that the perturbation expansion relevant to quantum theory is asymptotic in the sense of Poincaré. This provides a precise formulation and proof of Dyson's second conjecture[16] that, in general, we can only expect the expansion to be asymptotic.

Our investigation into the extent that our continuous models for the micro-world faithfully represent the amount of information available from experiment has led to a derivation of the time-ordered evolution operator in a more physical way. This approach made it possible to prove that the ultraviolet divergence is caused by a violation of the Heisenberg uncertainty relation at each point in time, thus partially confirming Dyson's first conjecture.

We used Dyson's original notation so as to explicitly exhibit the counter-term necessary to eliminate the self-energy divergence that occurs in QED.  This divergence is not accounted for and is outside the scope of the current investigation.  Thus, within our present framework, we cannot say that all the divergences arise from our disregard of some simple physics, and are not the result of deeper problems.  Thus, Dyson's concerns about the mathematical consistency of quantum electrodynamics, and quantum field theory in general, is still an open problem.

Although we are not working in the framework of axiomatic field theory, our approach may make some uneasy since Haag's theorem suggests that the interaction representation does not exist (see Streater and Wightman[27] pg. 161).  (Haag's theorem assumes, among other things, that the equal time commutation relations for the canonical variables of a interacting field are equivalent to those of a free field.)  In trying to explain this unfortunate result, these authors point out that ( see pg. 168) "… What is even more likely in physically interesting quantum field theories is that equal time commutation relations will make no sense at all; the field might not be an operator unless smeared in time as well as space. "  The work in Sections 5 and 6 of this paper strongly suggests that there is no physical basis to assume that we know anything about canonical variables at one instant in time (see postulate 2 and the following paragraph).  Thus, our approach actually confirms the above comments of Streater and Wightman.

**Acknowledgments**

Work for this paper was begun while the first author was supported as a member of the School of Mathematics in the Institute for Advanced Study, Princeton, NJ, and completed while visiting in the physics department of the University of Michigan.